\documentclass{article} 
\usepackage{nips15submit_e,times}
\usepackage{hyperref}
\usepackage{url}
\usepackage{amssymb}
\usepackage{amsthm}
\usepackage{braket}
\usepackage{amsmath}
\usepackage{subfigure}

\renewcommand{\vec}[1]{\mathbf{#1}}

\newcommand{\diag}{\operatorname{diag}}

\title{Efficient Computation of Decoherent Quantum Walks through Eigenvalue Perturbation}

\author{
Andrea Torsello \\
Universit\`{a} Ca' Foscari Venezia\\
\texttt{torsello@dais.unive.it} \\
\And
Luca Rossi \\
Aston University \\
\texttt{l.rossi@aston.ac.uk} \\
}

\nipsfinalcopy 

\begin{document}
\maketitle

\begin{abstract}
A number of recent studies have investigated the introduction of decoherence in quantum walks and the resulting transition to classical random walks. Interestingly, it has been shown that algorithmic properties of quantum walks with decoherence such as the spreading rate are sometimes better than their purely quantum counterparts. Not only quantum walks with decoherence provide a generalization of quantum walks that naturally encompasses both the quantum and classical case, but they also give rise to new and different probability distribution. The application of quantum walks with decoherence to large graphs is limited by the necessity of evolving a state vector whose size is quadratic in the number of nodes of the graph, as opposed to the linear state vector of the purely quantum (or classical) case. In this technical report, we show how to use perturbation theory to reduce the computational complexity of evolving a continuous-time quantum walk subject to decoherence. More specifically, given a graph over $n$ nodes, we show how to approximate the eigendecomposition of the $n^2 \times n^2$ Lindblad super-operator from the eigendecomposition of the $n \times n$ graph Hamiltonian.
\end{abstract}

\section{Introduction}
Quantum walks on graphs represent the quantum mechanical analogue of classical random walks~\cite{farhi1998quantum,aharonov2001quantum,kempe2003quantum}. Despite being similar in the definition, the dynamics of the two types of walks can be remarkably different, with quantum walks possessing a number of interesting properties not exhibited by their classical counterparts. In the classical case, the evolution of the walk is described by a real-valued probability vector. In the quantum case, the state is characterized by a complex-valued amplitude vector. An interesting consequence of this is that different paths are naturally allowed to destructively (constructively) interfere with each other.

Most of the work in the literature has considered pure quantum dynamics~\cite{farhi1998quantum,aharonov2001quantum,kempe2003quantum,childs2009universal,emms2009graph1,rossi2013characterizing,rossi2015measuring}, i.e, fully coherent quantum walks. However, it has been shown that the introduction of decoherence can result in some algorithmic properties of the walk, such as the spreading rate, being better than in the purely quantum case case~\cite{kendon2007decoherence,whitfield2010quantum}. Most importantly, quantum walks with decoherence represent a generalization of quantum walks that encompasses both classical and quantum walks, as well as new types of walks that result in different probability distributions~\cite{whitfield2010quantum}.

Recall that decoherence is the process by which a quantum system is altered by its interaction with the environment. The result of this process is a transition of the system from quantum to classical. For example, a quantum walk subject to decoherence transitions to a classical random walk, with a speed that depends on the decoherence rate. Unfortunately, while in the fully classical and fully quantum cases the size of the state vector is $n$, where $n$ denotes the number of nodes of the graph, in the decoherent case the size of the state vector is $n^2$. The Hamiltonian operator acting on the state vector of a unitary quantum walk is represented by a $n \times n$ matrix. With the addition of decoherence, on the other hand, the evolution is defined by the Lindblad super-operator, which is represented by a $n^2 \times n^2$ matrix. This clearly limits the possibility of analysing large graph structures using decoherent quantum walks.

In this technical report we propose to use perturbation theory~\cite{trefethen1997numerical,van2007computation} to reduce the computational complexity of evolving a continuous-time quantum walk subject to decoherence. In Section~\ref{quantumwalk} we introduce the necessary quantum mechanical background. In Section~\ref{perturbation} we review the eigenvalue perturbation problem and in Section~\ref{decoherence} we show how this can be applied to the problem at hand.

\section{Continuous-time Quantum Walks with Decoherence}\label{quantumwalk}
\subsection{Continuous-Time Quantum Walks}
The continuous-time quantum walk is the quantum analogous of the continuous-time random walk~\cite{farhi1998quantum}. Let $G = (V,E)$ denote an undirected graph with $n$ nodes. If $\vec{p}(t) \in \mathbb{R}^n$ denotes the state of walk at time $t$, in a continuous-time random walk the state vector evolves according to the equation $\vec{p}(t) = e^{-Lt} \vec{p}(0)$, where the graph Laplacian $L$ is the infinitesimal generator matrix of the underlying continuous-time Markov process.

Similarly to its classical counterpart, the state space of the continuous-time quantum walks is the vertex set of the graph. The classical state vector is replaced by a vector of complex amplitudes over $V$ whose squared norm sums to unity, and as such the state of the system is not constrained to lie in a probability space, thus allowing interference to take place. The general state of the walk at time $t$ is a complex linear combination of the basis states $\Ket{u}$, i.e.,
\begin{equation}
\Ket{\psi(t)} = \sum_{u\in V} \alpha_u(t) \Ket{u},
\end{equation}
where the amplitude $\alpha_u(t) \in \mathbb{C}$ and $\Ket{\psi(t)} \in \mathbb{C}^{|V|}$ are both complex. Moreover, we have that $\alpha_u(t) \alpha_u^*(t)$ gives the probability that at time $t$ the walker is at the vertex $u$, and thus $\sum_{u \in V} \alpha_u(t) \alpha^{*}_u(t) = 1$ and $\alpha_u(t) \alpha^{*}_u(t) \in [0,1]$, for all $u \in V$, $t \in \mathbb{R}^{+}$.

The evolution of the walk is governed by the Schr\"{o}dinger equation
\begin{equation}
\frac{\partial}{\partial t}\Ket{\psi(t)} = -iH\Ket{\psi(t)},
\end{equation}
where we denote the time-independent Hamiltonian as $H$. Generally speaking, a continuous-time quantum walk is induced whenever the structure of the graphs is reflected by the (0,1) pattern of the Hamiltonian. For example, we could take the adjacency matrix or the Laplacian. In the following we assume $H=L$.

Given an initial state $\Ket{\psi(0)}$, solving the Schr\"{o}dinger equation gives the expression of the state vector at time $t$,
\begin{equation}
\Ket{\psi(t)} = e^{-iLt}\Ket{\psi(0)}.
\end{equation}
This can be conveniently expressed in terms of the spectral decomposition of the Hamiltonian $H = \Phi \Lambda \Phi^\top$, i.e., $\Ket{\psi(t)} = \Phi^\top e^{-i \Lambda t} \Phi \Ket{\psi(0)}$,
where $\Phi$ denotes the $n \times n$ matrix $\Phi=(\phi_1 | \phi_2 | ... | \phi_j | ... | \phi_n)$ with the ordered eigenvectors $\phi_j$s of $H$ as columns and $\Lambda = \mbox{diag}(\lambda_1, \lambda_2, ..., \lambda_j, ..., \lambda_n)$ is the $n \times n$ diagonal matrix with the ordered eigenvalues $\lambda_j$ of $H$ as elements, and we have made use of the fact that $\mbox{exp}[-iLt]=\Phi^\top \mbox{exp}[-i \Lambda t] \Phi$.

\subsection{Quantum Walks with Decoherence}

The density matrix is introduced in quantum mechanics to describe a system whose state is an ensemble of pure quantum states $\Ket{\psi(i)}$, each with probability $p(i)$~\cite{nielsen2010quantum}. The density operator of such a system is defined as
\begin{equation}
\rho = \sum_i p(i) \Ket{\psi(i)}\Bra{\psi(i)}\,.
\end{equation}
For a quantum walk with state vector $\Ket{\psi(t)}$, the corresponding density matrix at time $t$ is $\rho(t) = \Braket{\psi(t) | \psi(t)}$. Similarly to the Schr\"{o}dinger equation, the Liouville-von Neumann equation describes how a density operator evolves in time
\begin{equation}
\frac{\partial}{\partial t} \rho(t) = -i[L,\rho(t)],
\end{equation}
where $L$ is the graph Laplacian and $[A, B] = AB - BA$ denotes the commutator.

We can add non-unitary decoherence to the system by writing~\cite{kendon2007decoherence}
\begin{equation}
\frac{\partial}{\partial t} \rho(t) = -i[L,\rho(t)] -p\rho(t)+p\mathcal{P}\rho(t),
\end{equation}
where $p$ is the rate per unit time with which we add decoherence to the walk, and $\mathcal{P}\rho(t) = \sum_j P_j \rho(t) P_j^{\dag}$ represents the effect of noise on $\rho(t)$, where $\lbrace P_j \rbrace$ is a set of projectors. Intuitively, the effect of the extra terms is to reduce the off-diagonal elements of $\rho(t)$, i.e., the coherence terms, at a rate $p$ per unit time~\cite{kendon2007decoherence}, while leaving the diagonal elements unaffected. More specifically, let $\operatorname{vec}\big(\rho(t)\big)$ be the vectorization of the density matrix $\rho(t)$. Then we can write
\begin{equation}\label{decoherence_eq}
\frac{\partial}{\partial t}\operatorname{vec}\big(\rho(t)\big) = \left[-i \left(L\otimes I + I\otimes-L\right) + p\left( \sum_{v\in V} E_{vv}\otimes E_{vv} - I\otimes I\right)\right] \operatorname{vec}\big(\rho(t)\big),
\end{equation}
where $E_{vv}$ is the matrix which is 1 in $(v,v)$ and 0 elsewhere, i.e., the projector on the node $v$.

\section{Eigenvalue Perturbation}\label{perturbation}

Let $A(t)$ be a $n\times n$ complex matrix parametrized by $t \in T\subseteq\mathbb{R}$. Further, assume that  $A(t)$ is diagonalizable for all values of $t$, i.e., there exist two $n\times n$ parametric matrices $X(t)$, $Y(t)$, and a diagonal  $n\times n$ matrix $\Lambda(t)$ such that for all $t\in T$
\begin{eqnarray}
 A(t) X(t) &=& X(t)\Lambda(t)\label{eq:right}\\
 Y(t) A(t) &=& \Lambda(t)Y(t)\label{eq:left}.
\end{eqnarray}

Without lack of generality, assume
\begin{eqnarray}
Y(t)X(t) &=& I\label{eq:inv}\\
\diag\big(X(t)^\dagger X(t)\big) &=& \mathbf{1}\label{eq:rvecnorm}.
\end{eqnarray}

We want to reconstruct $X(t)$, $Y(t)$, and $\Lambda(t)$ to the first order: 
\begin{eqnarray}
X(t) &\approx& X + t X^\prime\\
Y(t) &\approx& Y + t Y^\prime\\
\Lambda(t) &\approx&  \Lambda + t \Lambda^\prime,
\end{eqnarray}
where $X$, $X^\prime$, $Y$, $Y^\prime$, $\Lambda$, and $\Lambda^\prime$ are computed at time $t=0$.

To this end, since $A(t)$ is diagonalizable, $X$ and $Y$ have full rank, so we can write
\begin{eqnarray}
X^\prime = X B\\
Y^\prime = C Y,
\end{eqnarray}
for some matrices $B$ and $C$.

\subsection{Distinct Eigenvalues}

In the case where all the eigenvalues are distinct one can compute the eigenvalue and eigenvector derivatives directly. Differentiating (\ref{eq:right}), we have
\begin{equation}\label{eq:diff}
A^\prime X + A X^\prime = X^\prime \Lambda + X \Lambda^\prime.
\end{equation}
Left-multiplying both sides by $Y$ and recalling that $YAX=\Lambda$, we have
\begin{equation}\label{eq:diff2}
Y A^\prime X + \Lambda B = B\Lambda + \Lambda^\prime,
\end{equation}
from which
\begin{equation}
\diag(\Lambda^\prime) = \diag(Y A^\prime X) + \diag(\Lambda B - B\Lambda) = \diag(Y A^\prime X),
\end{equation}
from which we have the eigenvalue derivatives
\begin{equation}\label{eq:valdiff}
 \lambda^\prime_i = y_i^\dagger A^\prime x_i,
\end{equation}
where $x_i$ and $y_i$ are respectively the right and left eigenvectors corresponding to eigenvalue $\lambda_i$ of $A(0)$.

For the Eigenvectors, from (\ref{eq:diff2}) we have, for $i\neq j$
\begin{eqnarray}
(Y A^\prime X)_{ij} + (\Lambda B)_{ij} &=& (B\Lambda)_{ij} + (\Lambda^\prime)_{ij}\\
y_i^\dagger A^\prime x_j + \lambda_i b_{ij} &=& b_{ij}\lambda_j + 0,
\end{eqnarray}
from which
\begin{equation}\label{eq:mix}
 b_{ij} = \frac{y_i^\dagger A^\prime x_j}{\lambda_j-\lambda_i}.
\end{equation}

Differentiating (\ref{eq:rvecnorm}), we obtain
\begin{eqnarray}
 \mathbf{0} &=& \diag\big(X(t)^\dagger X(t)\big)^\prime = \diag\big((X^\prime)^\dagger X + X^\dagger X^\prime\big) \\\nonumber
 &=& \diag(B^\dagger X^\dagger X +  X^\dagger X B) = 2 \operatorname{Re}\big(\diag( X^\dagger X B ) \big)
\end{eqnarray}
so, for all $i$, we have
\begin{equation}
 \sum_k \operatorname{Re}\big((X^\dagger X)_{ik} b_{ki}\big) = 0\,,
\end{equation}
or, extracting the term for $k=i$, and recalling that $(X^\dagger X)_{ii}$=1
\begin{equation}
 \operatorname{Re}(b_{ii})= -\sum_{k\neq i} \operatorname{Re}\big( (X^\dagger X)_{ik} b_{ki}\big) 
\end{equation}
As for the imaginary part of $b_{ii}$, recall that even after the normalization constraint (\ref{eq:rvecnorm}) there is still a degree of freedom in the choice of the global phase of the eigenvectors $x_i$ which is reflected in an arbitrariness in the choice of $\operatorname{Im}(b_{ii})$. Here we set $\operatorname{Im}(b_{ii})=0.$

Note, also, that differentiating (\ref{eq:inv}) we obtain
\begin{equation}
 \mathbf{0} = \big(Y(t)X(t)\big)^\prime = Y^\prime X + Y X^\prime = CYX + YXB = C+B,
\end{equation}
from which we obtain
\begin{equation}
C=-B.
\end{equation}

\subsection{Repeated Eigenvalues}

In the case of repeated eigenvalues we have an additional degree of freedom from the choice of the eigenbasis. Any linear combination of eigenvectors corresponding to the same eigenvalue is still an eigenvector of the matrix, thus the observed eigenvectors $x_i$ can indeed be linear combinations of the limiting eigenvectors of $A(t)$ as $t \rightarrow 0$, resulting in a discontinuity. This can be solved by assuming that there is an unknown eigenvector basis $X$ that is continuous in $t$ and expressing it in terms of the observed eigenvector matrix $\hat{X}$:
\begin{equation}
X = \hat{X}\Gamma.
\end{equation}

Substituting into (\ref{eq:right}) and left-multiplying by $\hat{Y}=\hat{X}^{-1}$ we have
\begin{eqnarray}
 \hat{Y} A \hat{X}\Gamma &=& \hat{X}\Gamma\Lambda\\
 \Lambda \Gamma &=& \Gamma \Lambda,
\end{eqnarray}
Hence, $\Gamma$ is co-diagonalizable with $\Lambda$. Recall that $\Lambda$ is diagonal, thus $\Gamma$ must be block diagonal with the blocks corresponding to the repeated values of $\Lambda$.

Let $\bar{\lambda}$ be one such repeated eigenvalue, repeated with multiplicity $r$. We can partition the eigenvalue/eigenvector matrices as follows:
\begin{eqnarray}
&\Lambda = \left(\begin{array}{c c}\Lambda_1&\mathbf{0}\\\mathbf{0}&\bar{\lambda}I\end{array}\right) \qquad
\Gamma = \left(\begin{array}{c c}\Gamma_1&\mathbf{0}\\\mathbf{0}&\Gamma_2\end{array}\right) \qquad
B = \left(\begin{array}{c c}B_{11}&B_{12}\\B_{21}&B_{22}\end{array}\right)&\\
&\hat{X} = \left(X_1\; X_2\right) \qquad
\hat{Y} = \left(\begin{array}{c}Y_1\\Y_2\end{array}\right) \qquad&.
\end{eqnarray}

From (\ref{eq:diff2}) we obtain
\begin{equation}
\left(\begin{array}{c c}
Y_1 A^\prime X_1\Gamma_1 & Y_1 A^\prime X_2\Gamma_2\\
Y_2 A^\prime X_1\Gamma_1 & Y_2 A^\prime X_2\Gamma_2
\end{array}\right) 
=
\left(\begin{array}{c c}
\Gamma_1 (B_{11} \Lambda_1 - \Lambda_1 B_{11} - \Lambda_1^\prime) &
\Gamma_1 (\bar{\lambda}I-\Lambda_1)B_{12} \\
-\Gamma_2 B_{21} (\bar{\lambda}I-\Lambda_1) &
\Gamma_2 \Lambda_2^\prime
\end{array}\right).
\end{equation}

Hence, the block-diagonal element $\Gamma_2$ can be obtained by solving the eigenvalue problem
\begin{equation}\label{eq:eig2}
 Y_2 A^\prime X_2\Gamma_2 = \Gamma_2 \Lambda_2^\prime,
\end{equation}
where the derivatives $\Lambda_2^\prime$ of the repeated eigenvalues $\bar{\lambda}$ are the eigenvalues of $Y_2 A^\prime X_2\Gamma_2$. If these eigenvalues are distinct the matrix $\Gamma_2$ is unique up to a multiplicative factor.

If we assume that both the observed and the continuous eigenvectors are normalized, i.e.,
\begin{equation}
 \diag\big(X(t)^\dagger X(t)\big) = \mathbf{1} \qquad \diag\big(\Gamma^\dagger X(t)^\dagger X(t)\Gamma\big) = \mathbf{1},
\end{equation}
it is clear that on non-repeated eigenvalues, the corresponding diagonal element of $\Gamma$ must have norm 1. As usual the phase remains arbitrary, but we can pick $\gamma_{ii}=1$ without loss of generality.

The equation
\begin{equation}
 Y_1 A^\prime X_1\Gamma_1  = \Gamma_1 (B_{11} \Lambda_1 - \Lambda_1 B_{11} - \Lambda_1^\prime)
\end{equation}
is equivalent to (\ref{eq:diff2}) on the reduced eigenvalue set and can be recursively partitioned if $\Lambda_1$ still contains repeated eigenvalues and solved as in the case of non-repeated eigenvalues. On the other hand, the values of $B_{12}$ and $B_{21}$ can be computed from the following equations
\begin{eqnarray}
 Y_1 A^\prime X_2\Gamma_2 &=& \Gamma_1 (\bar{\lambda}I-\Lambda_1)B_{12}\label{eq38} \\
 Y_2 A^\prime X_1\Gamma_1 &=& -\Gamma_2 B_{21} (\bar{\lambda}I-\Lambda_1)\label{eq39}.
\end{eqnarray}

To compute $B_{22}$, we differentiate (\ref{eq:diff}) one more time, setting $X^{\prime\prime}=XD$, we left-multiply by $Y$ and we concentrate on the sub-matrix corresponding to the repeated eigenvalues:
\begin{equation}
 Y_2 A^{\prime\prime} X_2 \Gamma_2 + 2 Y_2 A^\prime \Gamma_2 B_{22} + \bar{\lambda}\Gamma_2 D = \Gamma_2 D \bar{\lambda} + 2 \Gamma_2 B_{22} \Lambda_2^\prime + \gamma_2\Lambda_2^{\prime\prime}.
\end{equation}
Recalling that $\Gamma_2$ is a solution to the eigenvalue problem (\ref{eq:eig2}), we have
\begin{equation}
  Y_2 A^{\prime\prime} X_2 \Gamma_2 + 2 \Gamma_2 \Lambda_2^\prime B_{22} =  2 \Gamma_2 B_{22} \Lambda_2^\prime + \gamma_2\Lambda_2^{\prime\prime}
\end{equation}
or
\begin{equation}
 2(B_{22} \Lambda_2^\prime - \Lambda_2^\prime B_{22} ) = \Gamma_2^{-1} Y_2 A^{\prime\prime} X_2 \Gamma_2  - \Lambda_2^{\prime\prime},
\end{equation}
from which we can extract the off-diagonal elements of $B_{22}$:
\begin{equation}\label{eq43}
 (B_{22})_{ij} = \frac{\Gamma_2^{-1} Y_2 A^{\prime\prime} X_2 \Gamma_2}{2(\bar{\lambda}_j^\prime-\bar{\lambda}_i^\prime)}.
\end{equation}
Note that in the special case where the matrix $A(t)$ is a linear function of $t$, {\em i.e.},  $A(t) = A + t A^\prime$, then $A^{\prime\prime}=\mathbf{0}$ and thus $(B_{22})_{ij} = 0$.

As for the non-repeated eigenvalue case, the diagonal of $B_{22}$ is computed from the constraint 
\begin{equation}
 \diag\big(\Gamma^\dagger X(t)^\dagger X(t)\Gamma\big) = \mathbf{1},
\end{equation}
resulting in 
\begin{equation}
 \operatorname{Re}(b_{ii})= -\sum_{k\neq i} \operatorname{Re}\big( (\Gamma^\dagger X^\dagger X \Gamma)_{ik} b_{ki}\big) 
\end{equation}
and, without loss of generality $\operatorname{Im}(b_{ii})=0.$

\subsection{Hermitian Matrices}
If $A$ is Hermitian, then $X$ is unitary and $Y=X^\dagger$. With this in mind, in the distinct eigenvalue case, we have
\begin{eqnarray}
 \lambda_i^\prime &=& x_i^\dagger A^\prime x_i\\
 b_{ij} &=& \frac{x_i^\dagger A^\prime x_j}{\lambda_j-\lambda_i}\\
 b_{ii} &=& 0.
\end{eqnarray}
Thus, if also $A^\prime$ is Hermitian, $B$ is skew-symmetric.

\section{Application to Quantum Walks with Decoherence}\label{decoherence}

Recall that the evolution of a quantum walk with decoherence expressed in terms of the density matrix $\rho$ is given by Eq.~\ref{decoherence_eq}. In order to compute the evolution of $\rho(t)$, we analyze the behavior of the eigenvalues and eigenvectors of 
\begin{equation}
A(p) = -i \left(L\otimes I + I\otimes-L\right) + p\left( \sum_{v\in V} E_{vv}\otimes E_{vv} - I\otimes I\right) = iA+pA^\prime
\end{equation}
as a function of the decoherence rate $p$.

For $p=0$ the eigenvalues of $A(p)$ are all of the form 
\begin{equation}
\pi_{jk} = i(\lambda_k - \lambda_j),
\end{equation}
where $\lambda_k$ and $\lambda_j$ are eigenvalues of $L$. The corresponding eigenvectors are of the form 
\begin{equation}
\xi_{jk} = \phi_j\otimes\phi_k,
\end{equation}
where $\phi_j$ is an eigenvector of $L$ corresponding to $\lambda_j$ and $\phi_k$ is an eigenvector  corresponding to $\lambda_k$.

Note that there is at least one  repeated eigenvalue in $A(0)$, namely $0$ with multiplicity at least $n$. In fact, for all $i=1,\ldots,n$ we have $\pi_{jj}=i(\lambda_j - \lambda_j)=0$ which is an eigenvalue with eigenvector $\xi_{jj}=\phi_j\otimes\phi_j$. 
In the following we make the simplifying assumption that this is the only case of repeated eigenvalue, namely that the eigenvalue gaps $\lambda_j - \lambda_k$  in $L$ are all unique for $j\neq k$.

Using (\ref{eq:valdiff}) we can compute the eigenvalue derivatives $\pi_{jk}\prime$ for $j\neq k$:
\begin{eqnarray}
\pi_{jk}\prime &=& \xi_{jk}^T A^\prime \xi_{jk} = (\phi_j\otimes\phi_k)^T \left( \sum_{v\in V} E_{vv}\otimes E_{vv} - I\otimes I\right) (\phi_j\otimes\phi_k) \nonumber\\
&=& \sum_{v\in V} (\phi_j^T E_{vv}\phi_j)\otimes (\phi_k^T E_{vv} \phi_k) - 1 = -\left(1-\sum_{v\in V} \phi_{jv}^2 \phi_{kv}^2 \right),
\end{eqnarray}
where the quantity 
\begin{equation}
o_{jk}=\sum_{v\in V} \phi_{jv}^2 \phi_{kv}^2
\end{equation}
is the probability of co-observation of the standing waves $\phi_j$ and $\phi_k$. This means that the (real) decay of the mixed eigenvector $\xi_{jk}$ introduced by the decoherence is proportional to the probability that the two components $\phi_j$ and $\phi_k$ are not observed on the same node.

For the eigenvector derivative, we compute the mixing proportion $b_{jk}^{lm}$ for $j\neq k$, $l\neq m$, and $(j,k)\neq(l,m)$. Intuitively, this tells us how much of $\xi_{jk}$ goes into $\xi_{lm}^\prime$
\begin{eqnarray}\label{eq54}
 b_{jk}^{lm} &=& \frac{\xi_{jk}^T A^\prime \xi_{lm}}{\pi_{lm}-\pi_{jk}} = \frac{(\phi_j\otimes\phi_k)^T \left( \sum_{v\in V} E_{vv}\otimes E_{vv} - I\otimes I\right) (\phi_l\otimes\phi_m)}{i(\lambda_m + \lambda_j - \lambda_l - \lambda_k)}\nonumber\\
 &=& \frac{\sum_{v\in V} (\phi_j^T E_{vv}\phi_l)\otimes (\phi_k^T E_{vv} \phi_m)}{i(\lambda_m + \lambda_j - \lambda_l - \lambda_k)} = \frac{\sum_{v\in V} \phi_{jv}\phi_{lv}\phi_{kv}\phi_{mv}}{i(\lambda_m + \lambda_j - \lambda_l - \lambda_k)}.
\end{eqnarray}
Hence, the mixing is proportional to the probability of co-observation of the standing waves $\phi_j$, $\phi_k$, $\phi_l$, and $\phi_m$.

For $j=k$ we have repeated eigenvalues, so we need to solve the following eigensystem:
\begin{equation}\label{eq55}
 \Xi \Gamma = \Gamma \Lambda(0)^\prime,
\end{equation}
with $\Xi=(\xi_{jk})$
\begin{eqnarray}
\xi_{jk} =  \xi_{jj} A^\prime \xi_{kk} &=& (\phi_j\otimes\phi_j)^T \left( \sum_{v\in V} E_{vv}\otimes E_{vv} - I\otimes I\right) (\phi_k\otimes\phi_k) \nonumber\\
&=& \sum_{v\in V} \phi_{jv}^2\phi_{kv}^2 - \delta_{jk},
\end{eqnarray}
thus we have that $\Xi = O - I$ where $O$ is the matrix of co-observations of the standing waves.
It is easy to show that $O$ is doubly stochastic, in fact
\begin{eqnarray}
 \sum_j o_{jk} &=&   \sum_j \sum_{v\in V} \phi_{jv}^2\phi_{kv}^2 = \sum_{v\in V} \left( \sum_j \phi_{jv}^2\right) \phi_{kv}^2 = \sum_{v\in V}  \phi_{kv}^2 =1\\
  \sum_k o_{jk} &=&  \sum_k \sum_{v\in V} \phi_{jv}^2\phi_{kv}^2 = \sum_{v\in V}  \phi_{jv}^2 \left(\sum_k \phi_{kv}^2\right) = \sum_{v\in V}  \phi_{jv}^2 =1,
\end{eqnarray}
thus, $\Xi$ has all real negative eigenvalue with the exception of at least one zero eigenvalue corresponding to the steady state of $O$.

\subsection{Computational Complexity}
We conclude this technical report with some remarks on the computational complexity of the proposed approach. To this end, note that we first need to compute the eigendecomposition of the Laplacian matrix $L$, which has complexity $O(n^3)$, where $n$ is the number of nodes of the graph. Similarly, solving the eigensystem of Eq.~\ref{eq55} has complexity $O(n^3)$, where $\Xi$ is a real-valued symmetric matrix and $\Gamma$ is orthogonal.

The computation of $B_{12}$ in Eq.~\ref{eq38} requires inverting $\Gamma_1$, which in our case is the identity matrix, and a diagonal matrix, i.e., $(\bar{\lambda}I-\Lambda_1)$. Similarly, solving Eq.~\ref{eq39} for $B_{21}$ requires inverting $(\bar{\lambda}I-\Lambda_1)$ and $\Gamma_2$. Since $\Gamma$ is orthogonal and block-diagonal, we conclude that $\Gamma_2$ is an orthogonal matrix. In general, note that $B$ is an $n^2 \times n^2$ matrix and therefore the complexity of constructing it is at least $O(n^4)$. In particular, from Eq.~\ref{eq54} it follows that the complexity of computing the $n^4$ elements of $B$ is $O(n^5)$. We should stress, however, that the computation of the $b_{jk}^{lm}$ can be easily parallelized. 

As a result, we conclude that the complexity of the proposed approach is dominated by the $O(n^5)$ computation of the matrix $B$. This should be contrasted with the cost of directly computing the eigendecomposition of the $n^2 \times n^2$ super-operator $A(p)$, which is $O(n^6)$. Finally, note that for a generic $p > 0$, $A(p)$ is not Hermitian and therefore techniques like singular value decomposition cannot be employed.

\bibliography{biblio}
\bibliographystyle{unsrt}

\end{document}